\pdfoutput=1

\documentclass[aps,prl,reprint,superscriptaddress,floatfix,nofootinbib,longbibliography]{revtex4-2}

\usepackage{graphicx}
\usepackage{amsmath, amssymb, mathrsfs}
\usepackage{bm}
\usepackage[T1]{fontenc}
\usepackage[utf8]{inputenc}
\usepackage{microtype}
\usepackage[dvipsnames]{xcolor}
\usepackage[colorlinks,urlcolor=NavyBlue,citecolor=NavyBlue,linkcolor=NavyBlue,pdfusetitle]{hyperref}
\usepackage[all]{hypcap}
\usepackage{orcidlink}
\usepackage{tikz}
\usetikzlibrary{decorations.pathmorphing, arrows.meta, calc}

\graphicspath{{Figures}}

\newcommand{\Igroup}{\mathscr{I}^{+}}

\newsavebox{\prlSchematicBox}
\newcommand{\CornellPhysics}{\affiliation{Department of Physics, Cornell University, Ithaca, NY 14853, USA}}
\newcommand{\Cornell}{\affiliation{Cornell Center for Astrophysics and Planetary Science, Cornell University, Ithaca, NY 14853, USA}}
\newcommand{\CornellLepp}{\affiliation{Laboratory for Elementary Particle Physics, Cornell University, Ithaca, NY 14853, USA}}
\newcommand{\Caltech}{\affiliation{Theoretical Astrophysics 350-17, California Institute of Technology, Pasadena, CA 91125, USA}}
\newcommand{\UMiss}{\affiliation{Department of Physics and Astronomy, University of Mississippi, University, MS 38677, USA}}

\newcommand{\figMain}{%
\begin{figure*}[t]
    \begin{center}
    \includegraphics[width=\textwidth]{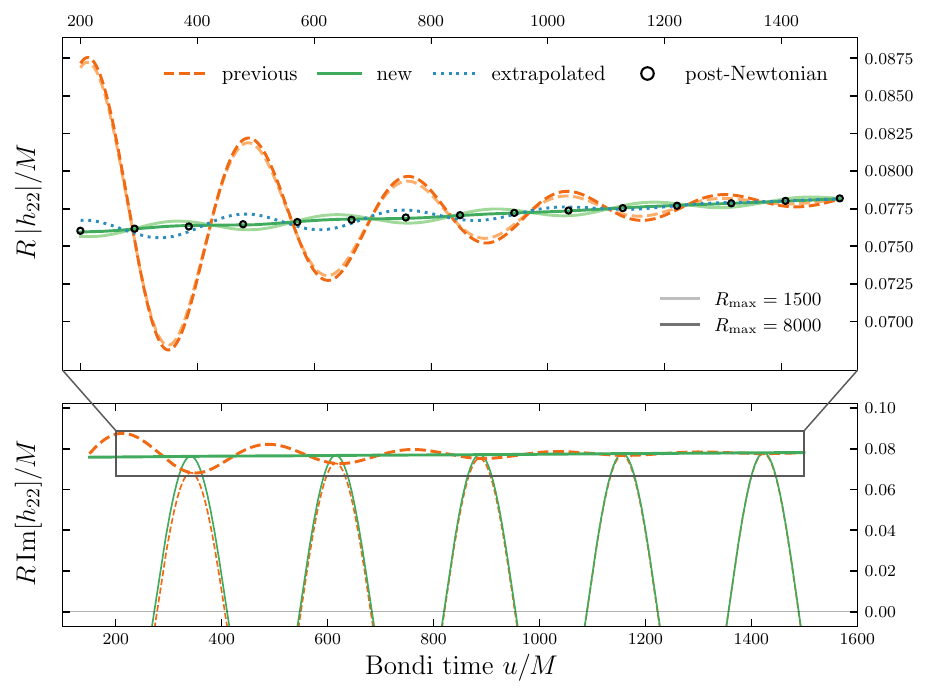}
    \end{center}
    \caption{The $(2,2)$-mode of the strain obtained with the
      \texttt{SpECTRE} CCE module for an equal-mass, non-spinning,
      quasi-circular BBH. Top: amplitude of the $(2,2)$-mode for the previous
      \emph{ConformalFactor} initial data (orange, dashed) and the new
      \emph{CauchySecondOrder} scheme (green, solid) for two outer-boundary radii
      $R_{\mathrm{max}}=1500,\,8000\,M$ (light, dark), a
      post-Newtonian prediction (black circles), and an extrapolated waveform~\cite{Boyle:2009vi,Iozzo:2020jcu} (blue, dotted) with outer-boundary radius $R_{\mathrm{max}}=1500\,M$.
      Bottom: imaginary part of the $(2,2)$-mode over a longer window of the
      simulation (itself a zoom of the full evolution), with the boxed
      region showing where the top panel zooms in further. With the
      previous scheme the amplitude oscillates by
      up to $\sim10\%$ of its value for over $1000\,M$; with the new
      scheme it tracks the post-Newtonian prediction from the initial slice.}
    \label{fig:main}
\end{figure*}
}

\begin{document}

\title{Calculating gravitational waves at infinity: the missing link}

\author{Guido Da Re \orcidlink{0009-0007-2006-094X}} \email{gdare@caltech.edu} \Caltech
\author{Keefe Mitman \orcidlink{0000-0003-0276-3856}} \email{kem343@cornell.edu} \Cornell
\author{Leo C. Stein \orcidlink{0000-0001-7559-9597}} \UMiss
\author{Nils Deppe \orcidlink{0000-0003-4557-4115}} \CornellLepp \CornellPhysics \Cornell
\author{Himanshu Chaudhary \orcidlink{0000-0002-4101-0534}} \Caltech
\author{Dongze Sun \orcidlink{0000-0003-0167-4392}} \Caltech
\author{Michael Boyle \orcidlink{0000-0002-5075-5116}} \Cornell
\author{Kyle C.~Nelli \orcidlink{0000-0003-2426-8768}} \Caltech
\author{Mark A. Scheel \orcidlink{0000-0001-6656-9134}} \Caltech
\author{Saul A. Teukolsky~\orcidlink{0000-0001-9765-4526}} \Caltech\Cornell

\date{\today}

\begin{abstract}
    Gravitational-wave science requires waveform templates to be
    computed at future null infinity, the asymptotic boundary of
    spacetime. Numerical solutions of Einstein's equations typically
    rely on a Cauchy evolution on
    a finite computational domain, but can produce waveforms
    at infinity by performing a subsequent characteristic evolution
    that compactifies the spacetime. Such waveforms capture
    expected physical effects, like gravitational-wave memory and
    late-time, power-law tails. But this characteristic evolution requires initial
    data consistent with the interior Cauchy solution
    to avoid producing spurious, long-lived transients.
    A careful analysis shows that previous prescriptions for this
    initial data were not sufficiently consistent and provides
    a way to remedy the problem.
    The new initial data yield accurate waveforms from the
    first cycle onward.  The construction is independent
    of the Cauchy formulation and enables the production of
    catalogs of robust waveforms at infinity.
\end{abstract}

\maketitle

\textit{Introduction.---} Many of the radiative observables in astronomy and astrophysics are measured not near their sources but essentially infinitely far away. This necessity is also a virtue: radiation only obtains an unambiguous physical interpretation at infinity. In any gauge theory, including both electromagnetism and general relativity, the radiative degrees of freedom cleanly separate from the Coulombic ones as an observer approaches infinity, because of their distinct radial falloffs~\cite{Bondi:1962px,Sachs:1962wk}.\footnote{For general relativity the separation is unique only up to the global BMS (Bondi--Metzner--Van der Burg--Sachs) symmetries of future null infinity, under which the radiation transforms covariantly.} In the bulk, however, local gauge freedom renders this split ambiguous.

In 2015, GW150914 marked the first direct detection of gravitational waves
(GWs) from a binary black hole (BBH) merger~\cite{LIGOScientific:2016aoc}.
Two years later, GW170817 revealed a binary neutron-star merger with an
electromagnetic counterpart~\cite{LIGOScientific:2017adf}.
As the sensitivity of the
LIGO-Virgo-KAGRA (LVK) detectors improves,
observations with increasingly high signal-to-noise ratios are expected. The binary black hole merger GW250114, for example, enabled the most stringent tests of general
relativity (GR) to date~\cite{LIGOScientific:2025rid,LIGOScientific:2025wao}. Exploiting
the information contained in such observations requires waveform models accurate
enough to support precise parameter inference~\cite{Purrer:2019jcp}, tests of GR~\cite{LIGOScientific:2025rid,LIGOScientific:2025wao}, and standard-siren
cosmology~\cite{LIGOScientific:2017adf}.

Numerical relativity (NR) simulations provide the only means of solving the full
nonlinear Einstein equations for a binary black hole
merger~\cite{Pretorius:2005gq}. However, the waveforms used to train the LVK
Collaboration's models come from public NR
catalogs~\cite{Boyle:2019kee,Scheel:2025jct,Healy:2017psd,Healy:2019jyf,
Healy:2020vre,Healy:2022wdn,Ficarra:2026cej,Jani:2016wkt,Ferguson:2023vta,
Hamilton:2023qkv,Huerta:2019oxn} and are typically computed without evolving the
spacetime out to future null infinity. Instead, these waveforms are extrapolated
to infinity from finite-radius data~\cite{Boyle:2019kee},
either by polynomial fits in $1/r$~\cite{Boyle:2009vi,Iozzo:2020jcu} or by
perturbative extraction~\cite{Nakano:2015pta}. This extrapolation results in less accurate
waveforms that do not accurately exhibit expected physical effects, like
momentum fluxes and GW memory~\cite{DaRe:2025glj,Favata:2010zu}, which will be
increasingly important for future
detections~\cite{Varma:2020Recoil,Mitman:2026zfg}.

To accurately reach infinity, researchers instead use a \emph{characteristic evolution} that solves Einstein's equations on null hypersurfaces. By conformally compactifying the spacetime in this secondary evolution, future null infinity is included in the computational domain~\cite{Bishop:1996gt,Handmer:2016mls,Moxon:2020gha,Moxon:2021gbv}.\footnote{An alternative is hyperboloidal evolution, which uses spacelike slices
that approach future null infinity (see Refs.~\cite{Rinne:2009Hyperboloidal,Peterson:2024Hyperboloidal} and references therein).}
 This evolution occurs after the standard, finite-domain simulation performed on spacelike slices. Consequently, it is often called Cauchy-characteristic evolution (CCE), since the interior Cauchy simulation
supplies the inner boundary data for the exterior characteristic evolution.
Waveforms output by CCE have been shown to accurately capture certain physical effects that the more commonly used extrapolated waveforms do not~\cite{Pollney:2010hs,Mitman:2020pbt,DaRe:2025glj}. Hence, to ensure that analyses of GW observations are not biased by systematic errors in the computation of NR waveforms at null infinity, accurate evolution to infinity is vital for computing GR-consistent waveforms.

A critical current limitation in the accuracy of CCE, however, is the unphysical radiation that is sourced by inconsistent characteristic initial data. This results in a long-lived, low-frequency transient that pollutes several cycles of the inspiral and degrades post-ringdown features, like tails~\cite{DeAmicis}.
Such waveform errors can bias parameter estimation~\cite{Purrer:2019jcp}.
Therefore, as our detectors become more sensitive, eliminating this source of systematic error is essential for robust science.

In this \textit{Letter}, we present a method of generating characteristic
initial data that effectively eliminates spurious CCE transients in binary black hole simulations and recovers the correct late-time tail behavior in
head-on black hole collisions. We match the initial radial profile to the
Cauchy data through second order, enforce asymptotic regularity, and determine
the initial angular coordinates from that profile. Our new method,
which is independent of the Cauchy formulation, enables the production of NR
waveform catalogs using CCE that provide the five Weyl curvature components in
addition to the gravitational wave strain, all while capturing expected
physical features. This enables the construction of more accurate and physically
consistent waveform models. It allows the gravitational physics community to
study the merger and ringdown processes in unprecedented detail, ultimately
unlocking a new frontier in gravitational wave science.

\textit{The characteristic initial-data problem.---}
	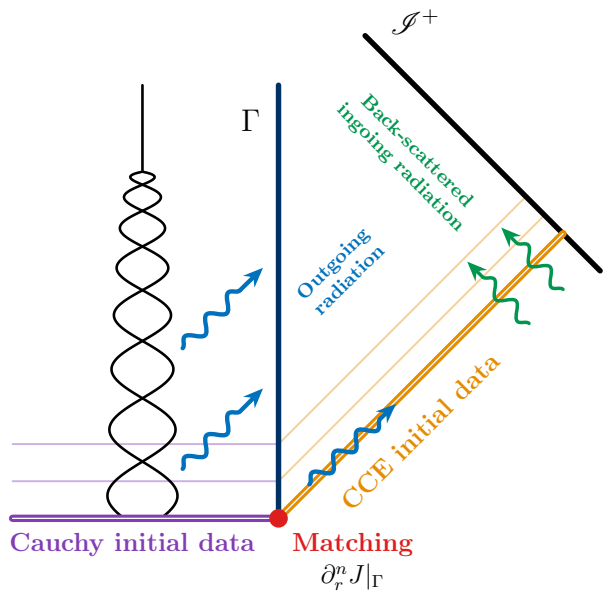
\begin{figure}[t]
		\centering

		\sbox{\prlSchematicBox}{%
\providecommand{\Igroup}{\mathscr{I}^{+}}

\definecolor{darkblue}{RGB}{0, 48, 100}
\definecolor{myblue}{RGB}{0, 114, 189}
\definecolor{mygreen}{RGB}{0, 150, 80}
\definecolor{myorange}{RGB}{230, 140, 0}
\definecolor{myred}{RGB}{220, 30, 30}
\definecolor{mypurple}{RGB}{130, 60, 180}

\begin{tikzpicture}[
    line cap=round,
    outgw/.style={line width=2.0pt, -{Stealth[length=3.2mm, width=2.6mm]},
      decorate, decoration={snake, amplitude=1.1mm, segment length=5mm,
        post length=3mm}},
    ingw/.style={line width=1.6pt, -{Stealth[length=3.2mm, width=2.6mm]},
      decorate, decoration={snake, amplitude=1.1mm, segment length=5mm,
        post length=3mm}},
    cauchyfoliation/.style={mypurple!40, line width=0.9pt},
    ccefoliation/.style={myorange!40, line width=0.9pt},
  ]

  
  \draw[cauchyfoliation] (-4.3,0.6) -- (0,0.6);
  \draw[ccefoliation]    (0,0.6) -- (4.3,4.9);
  
  \draw[cauchyfoliation] (-4.3,1.2) -- (0,1.2);
  \draw[ccefoliation]    (0,1.2) -- (4.0,5.2);

  \begin{scope}[shift={(-2.2,0)}]
    \newcommand{\Tmerg}{5.6}
    \newcommand{\Tscale}{30}
    \newcommand{\Rscale}{0.16}
    \draw[black, very thick, variable=\t, domain=0:\Tmerg, samples=500]
      plot ({\Rscale * (\Tscale*(\Tmerg-\t))^(1/4)
             * sin((\Tscale*(\Tmerg-\t))^(5/8) r)}, {\t});
    \draw[black, very thick, variable=\t, domain=0:\Tmerg, samples=500]
      plot ({-\Rscale * (\Tscale*(\Tmerg-\t))^(1/4)
             * sin((\Tscale*(\Tmerg-\t))^(5/8) r)}, {\t});
    \draw[black, very thick] (0,\Tmerg) -- (0,7.0);
  \end{scope}

  \draw[mypurple, line width=1.4pt, double] (-4.3,0) --
    node[pos=0.45, below=3pt, font=\large\bfseries] {Cauchy initial data}
    (0,0);

  \draw[myblue, outgw] (-1.55, 0.80) -- (-0.25, 2.10);
  \draw[myblue, outgw] (-1.55, 2.75) -- (-0.25, 4.05);

  \draw[darkblue, line width=2.5pt] (0,0) --
    node[pos=0.92, left=4pt, black, font=\Large] {$\Gamma$} (0,7);

  \draw[black, line width=2.5pt] (1.4,7.8) --
    node[pos=0.05, above right=2pt, black, font=\Large] {$\Igroup$}
    (5.2,4.0);

  \draw[myorange, line width=1.4pt, double] (0,0) -- (4.6,4.6);
  \node[myorange, font=\large\bfseries, rotate=45] at (2.29,1.41)
    {CCE initial data};


  \draw[myblue, outgw] ($(0.5,0.55)$) -- ($(1.9,1.85)$);
    
  \node[myblue, font=\small\bfseries,
  rotate=45,
  align=center] at (1.0,4.0) {Outgoing\\ radiation};

  \draw[mygreen, ingw] (4.05,3.15) -- (3.05,4.15);
  \draw[mygreen, ingw] (4.60,3.70) -- (3.60,4.70);
  \node[mygreen, font=\small\bfseries,
  rotate=-45,
  align=center] at (2.1,5.8)
    {Back-scattered\\ ingoing radiation};

  \fill[myred] (0,0) circle (0.15);
  \node[myred, anchor=north, font=\large\bfseries] at (1.2,-0.12) {Matching};
  \node[black, anchor=north, font=\large] at (1.2,-0.62)
    {$\partial_r^n J\rvert_\Gamma$};

\end{tikzpicture}%
		}%

		\resizebox{\columnwidth}{!}{\usebox{\prlSchematicBox}}%

		\caption{The characteristic initial-data problem. A Cauchy evolution
			(left) supplies data on a worldtube $\Gamma$; a characteristic evolution
			on null slices propagates radiation to future null infinity $\Igroup$. The initial data
			(orange slice), specified through the Bondi scalar $J$, encode outgoing
			(blue) and back-scattered ingoing (green) radiation from the entire causal
			past. However, the only available information is the worldtube data at the
			matching point (red) and the radial derivatives
			$\partial_r^n J\rvert_\Gamma$.}
		\label{fig:schematic}
	\end{figure}

In CCE, a Cauchy evolution supplies the spacetime metric and its
derivatives on a timelike worldtube $\Gamma$ of finite radius $R$. From
these data, an exterior Bondi-like coordinate system is constructed:
spacetime is foliated by a series of outgoing null hypersurfaces $\Sigma_u$ labeled
by retarded time $u$, spanned by null rays extending from $\Gamma$ to
future null infinity $\Igroup$ (Fig.~\ref{fig:schematic})~\cite{Bondi:1962px,Sachs:1962wk,
	Bishop:1996gt,Moxon:2020gha}. In this gauge, Einstein's equations
reduce on each null slice to a hierarchy of first-order radial ordinary
differential equations (the hypersurface equations) that can be integrated outward from
$\Gamma$ to $\Igroup$. The field that serves as the source for these equations is the complex spin-weight $+2$ scalar $J$: a component of the rescaled transverse metric
whose $1/r$ coefficient at $\Igroup$ is closely related to the gravitational wave strain in Bondi--Sachs coordinates.

Since $J$ supplies the source for
the characteristic evolution, it must be specified on the first null hypersurface $\Sigma_0$.
Unlike Cauchy initial data, which obey
elliptic constraint equations, the initial $J$ is not determined by a
radial constraint on a single null hypersurface.
Instead, the constraints take the form of flux-balance laws in retarded time,
leaving the initial $J$ freely specifiable.
Physically, however, it is anything but free: $J$ on $\Sigma_0$
encodes both outgoing radiation
and ingoing radiation back-scattered from the system's entire causal past.
This information is not fully available:
the Cauchy simulation provides the spacetime metric
neither outside $\Gamma$ nor before its own initial time.

Nonetheless, we can seek consistency between the initial Cauchy and
characteristic slices by requiring that they describe the same ingoing and outgoing radiation.
That consistency is carried by the radial derivatives $\partial_r^{n}J\rvert_\Gamma$, which fix the radial profile of
$J$ on $\Sigma_0$ order by order. Matching the differential structure of $J$ at the worldtube is
therefore a local proxy for initializing the ingoing and outgoing radiation on the whole null slice.

Past approaches, ranging from perturbative solutions to matching
conditions~\cite{Bishop:1996gt,Babiuc:2008qy,Barkett:2019uae}, established
that a stable evolution requires the characteristic initial data to be at least continuous
with the worldtube data~\cite{Moxon:2020gha}, and that accuracy demands
higher-order differentiability. Existing schemes therefore use a truncated analytic expansion
of $J$ in $1/r$ and fix its
coefficients by matching the worldtube value and first radial derivative.

The angular coordinates $\phi^A$ that label the outgoing null rays are the Cauchy simulation's own, carried unchanged along each ray to $\Igroup$. That gauge is chosen for stability in the strong-field region and is not asymptotically inertial: $J$ retains an angular gauge contribution that does not decay, i.e., $J \sim \mathcal{O}(1)$ rather than $\mathcal{O}(1/r)$. In this gauge, the constant and $r^{-2}$ contributions to $J$
can produce a net $1/r$ contribution to the hypersurface equations~\cite{Moxon:2020gha},
as they contain derivatives and nonlinear combinations of the fields
weighted by powers of $r$. Radial integration then generates
$r^{-n}\log r$ terms, degrading spectral convergence~\cite{Moxon:2020gha,Moxon:2021gbv}.
Therefore, during the evolution, the worldtube angular
coordinates inherited from the Cauchy
simulation are mapped through a time-dependent transformation to
``partially flat coordinates'' $\hat{\phi}^{\hat A}$, hereafter called the angular map.

In the partially flat gauge, the initial conditions
$\hat J^{(0)}=\hat J^{(2)}=0$ ensure that the net
$1/\hat r$ contributions to the radial integrands vanish, preventing
logarithmic terms. Here $\hat J^{(n)}$ denotes the coefficient of
$\hat r^{-n}$. These conditions are then preserved by the
evolution~\cite{Moxon:2020gha}.
The angular map is quasi-conformal~\cite{companion}: it splits uniquely into a conformal part, which the gauge condition leaves free, and a non-conformal part, which it fixes. The map
$\hat{\phi}^{\hat{A}}(u, \phi^A)$ is itself evolved during the simulation,
so its value at the initial time is part of the initial data.

The initial data thus comprise the radial profile $J$ on $\Sigma_0$,
the initial angular map $\hat{\phi}^{\hat A}(u_0,\phi^A)$, and the initial
Bondi time at $\Igroup$. The initial Bondi time and the residual conformal
freedom of the angular map select a BMS (Bondi--Metzner--Van der Burg--Sachs) frame.
Orientation-preserving conformal maps of the sphere form the Lorentz group
$\mathrm{SO}^{+}(1,3)$, while the choice of initial Bondi time fixes a
supertranslation.
We fix this gauge freedom in post-processing~\cite{Mitman:2021xkq,Mitman:2024uss}.

Existing schemes in the \texttt{SpECTRE} code~\cite{spectreCode} fix the initial angular map
$\hat{\phi}^{\hat{A}}(u_{0},\phi^A)$ by an a priori choice rather than deriving it from the worldtube data.\footnote{The one exception is the \emph{NoIncomingRadiation} scheme discussed below, which does derive the map, but from a radial profile of insufficient accuracy.} The most successful scheme, in that it reduces the spurious transients in the memory modes of quasi-circular binaries, is the
\emph{ConformalFactor} scheme, which postulates the map by requiring that the coordinate time at the worldtube agrees with the one measured by an asymptotic observer, a condition designed to bring the frame close to
Bondi gauge~\cite{Moxon:2020gha}. After the worldtube variables are transformed
into the new gauge, the radial ansatz is matched to first order, setting $\hat J^{(0)} = \hat J^{(2)} = 0$ by
construction. However, the resulting initial data describe a different spacetime from the one
supplied by the Cauchy simulation: the prescribed angular map is generally incompatible with the partially flat conditions for that spacetime.
The evolution radiates the difference away as spurious unphysical transients.

The mechanism creating this transient is clearest in a nonradiating spacetime. In the
Bondi--Sachs gauge, the Schwarzschild worldtube data have $J|_\Gamma=0$,
and the exact solution has $J=0$ everywhere on every null slice. Applying
an angular transformation throughout the slice gives a profile independent
of $\hat r$, which is nonzero for a nonconformal map~\cite{Moxon:2020gha,Moxon:2021gbv}.
Any physically equivalent initial data
within this coordinate family must likewise be independent of $\hat{r}$. The partially flat gauge
selects a subfamily of these maps through the condition $\hat J^{(0)}=0$.
For Schwarzschild, this requires a purely conformal map.

Now follow the existing schemes: first, apply a nonconformal angular map
at the worldtube. For Schwarzschild, this generates a nonzero $\hat{J}$. Then match the radial profile of $\hat J$, imposing
$\hat J^{(0)}=0$ directly on the series expansion. The resulting profile
decays as $\hat J=\mathcal{O}(1/\hat r)$ and describes a gauge-inequivalent
spacetime with unphysical radiation. Higher-order
matching with the same prescribed angular map
cannot resolve the problem because it preserves this falloff.
This inconsistency is precisely what powers the spurious radiation observed in all simulations.

The condition $\hat J^{(0)}=0$ thus constrains the initial angular
map, leaving only its conformal freedom. Unlike in the Schwarzschild
example, the selected map is typically nonconformal for a general
system such as a binary.
Its nonconformal part is fixed by the asymptotic
angular metric in the Cauchy-inherited coordinates, encoded in
$J^{(0)}$. Computing the map accurately therefore requires reconstructing
$J^{(0)}$ at $\Igroup$ from finite-radius worldtube data through
higher-order matching of the Cauchy-gauge radial profile, as we
describe below~\cite{companion}.

\textit{Characteristic junk in binary black holes.---}
\figMain
Figure~\ref{fig:main} shows the strain of the dominant $(2,2)$ mode at
$\Igroup$ for an equal-mass, non-spinning, quasi-circular BBH.
The Cauchy evolutions use \texttt{SpEC}~\cite{Boyle:2019kee} at \texttt{Lev3} resolution.
We obtain the strain with the \texttt{SpECTRE} CCE
module~\cite{Moxon:2020gha,Moxon:2021gbv} by evolving worldtube data
extracted at $R=300\,M$.
The worldtube data are stored through $\ell_{\max}=16$, and the
characteristic evolution uses $\ell_{\max}=20$ and $15$ radial collocation
points.
With the existing \emph{ConformalFactor} initial data, the amplitude
oscillates by up to $\sim$10\% about its expected value, and the
oscillation persists for more than $1000\,M$ of evolution.

We call this spurious transient the \emph{characteristic junk}, by
analogy with the Cauchy junk of the initial-value problem. The two are
distinguished by their origin, not by when they appear.  The Cauchy junk originates inside the Cauchy domain---in the relaxation of its initial data, and in the treatment of its finite outer boundary---and reaches $\Igroup$ through the worldtube data themselves~\cite{companion}.  The characteristic junk is sourced on the initial
null hypersurface itself, is of low frequency, and decays slowly. Because the
characteristic initial data are matched to the Cauchy data, it might be
tempting to conclude that the Cauchy initial data, with their imperfect
radiation content, are what trigger the characteristic junk.
However, restarting
the characteristic evolution after the
burst of Cauchy junk and its reflections have crossed the worldtube regenerates the characteristic junk in full for every existing
scheme~\cite{companion}. Its origin is not the Cauchy initial data, but
an inconsistency between the Cauchy and characteristic slices.

\textit{Cauchy-gauge matching.---}
The new initial-data scheme first constructs the radial profile of $J$ in the Cauchy gauge using the truncated asymptotic series
\begin{equation}
	J\big|_{\Sigma_0} = \sum_{n=0}^{N+1}\frac{J^{(n)}(\phi^{A})}{r^{n}}.
	\label{eq:ansatz}
\end{equation}
Matching the
value and radial derivatives of this series through order $N$ to those
supplied by the Cauchy evolution at the worldtube provides $N+1$ conditions
for the $N+2$ coefficients. The remaining asymptotic regularity condition
$\hat J^{(2)}=0$ fixes the remaining freedom through a nonlinear equation,
which we solve by fixed-point iteration.
At each fixed matching order, the companion paper in Ref.~\cite{companion} proves
existence and uniqueness within a small-amplitude region
and convergence of the fixed-point iteration under a bound satisfied by radiation-zone worldtube data.

With the Cauchy-gauge profile reconstructed, we now determine the
angular map selected by $\hat J^{(0)}=0$. This condition takes the form of a
Beltrami equation on $S^2$, whose coefficient $\mu$ is determined by the asymptotic
angular metric encoded in $J^{(0)}$~\cite{companion}:
\begin{equation}
	\mu=-\frac{J^{(0)}}{1+\sqrt{1+|J^{(0)}|^2}},
	  \qquad |\mu|<1.
	\label{eq:gaugecondition}
\end{equation}
Here $\mu$ measures the map's departure from conformality, with $\mu=0$ corresponding to a conformal map. For bounded $J^{(0)}$, the coefficient
satisfies $|\mu|<1$ uniformly on $S^2$, which guarantees a quasi-conformal solution, unique
up to conformal transformations~\cite{companion,AhlforsBers:1960}.
The angular map enforces $\hat J^{(0)}=0$, while the nonlinear constraint
imposed during radial matching ensures $\hat J^{(2)}=0$ in the transformed profile.

The angular map is thus fixed by the reconstructed $J^{(0)}$, whose
accuracy depends on the radiation content retained by the radial ansatz.
The matching order $N$ is therefore crucial, as it determines this accuracy.

At leading order in the field amplitude, $N=1$ reproduces the
\emph{NoIncomingRadiation} approximation, which sets $\Psi_0=0$~\cite{Moxon:2020gha,companion}.
That scheme already derives its angular map from the Cauchy-gauge radial
profile, but approximately neglects incoming radiation. Matching the second radial
derivative independently retains radial information omitted by this approximation,
improving the reconstructed $J^{(0)}$ and hence the angular map.

The second-order ($N=2$) version, \emph{CauchySecondOrder}, is our improved method; it is implemented in the open-source \texttt{SpECTRE} code~\cite{spectreCode} and is the focus of the following results.

The required second radial derivative is recovered from
the Bondi form of Einstein's equations at the worldtube using the stored
worldtube Cauchy data~\cite{companion}. All the existing simulations in the SXS
catalog~\mbox{\cite{Scheel:2025jct}}, as well as simulations using other
Cauchy formulations with stored worldtube data, can therefore be reprocessed
with the new scheme, without repeating their Cauchy evolutions.
This is possible because the \texttt{SpECTRE} worldtube
preprocessor~\cite{spectreCode} constructs the required Bondi--Sachs input from
either the spacetime metric of generalized-harmonic
evolutions~\cite{Pretorius:2005gq,Lindblom:2005qh} or the ADM variables of
moving-puncture evolutions~\cite{Campanelli:2005dd,Baker:2005vv}.

\textit{Numerical results.---}
Figure~\ref{fig:main} also shows the new second-order Cauchy-gauge
matching, together with a post-Newtonian prediction of the $(2,2)$
amplitude, for two placements of the Cauchy outer boundary. The oscillation
of the previous scheme is essentially independent of $R_{\mathrm{max}}$,
whereas with the new scheme the waveform tracks the post-Newtonian prediction from the
beginning.\footnote{The post-Newtonian waveform is generated with the
	\textsc{PostNewtonian.jl} package,
	\url{https://github.com/moble/PostNewtonian.jl}.}

The small residual modulation that remains is not characteristic junk.
It decreases systematically as the outer boundary is moved outward, and
is shared by the extrapolated waveform of the same system (at an
outer boundary radius of $R_{\mathrm{max}}=1500\,M$). We identify it as
Cauchy outer-boundary junk contaminating the worldtube data themselves. The
new initial data thus suppress the characteristic transient below the
outer-boundary contamination, and the error budget at $\Igroup$ is no
longer dominated by the characteristic initial data but by the accuracy of the worldtube data themselves: a separate problem that finds a natural solution in Cauchy-characteristic matching~\cite{Ma:2023qjn}.

The impact extends beyond the inspiral: for the
head-on collision studied in Ref.~\cite{DeAmicis}, the incorrect
late-time tail obtained with the previous initial data is fully
corrected by the new scheme~\cite{companion}, recovering the behavior of the most
reliable predictions in the literature~\cite{Ma:2024qcv,DeAmicis}.

\textit{Discussion.---} Spurious radiation
from inconsistent
characteristic initial data of CCE simulations has been the primary factor
restricting the release of numerical relativity waveform catalogs with CCE. In
this \textit{Letter}, we address this issue by constructing the initial
angular map from a radial profile reconstructed in the Cauchy gauge.
Matching through second order retains radial information omitted by the no-incoming-radiation approximation.
The resulting scheme suppresses the
characteristic transient below the worldtube contamination in the binary
studied here, and recovers the correct late-time tail in a head-on
collision~\cite{companion}.

The scheme allows archived simulations~\cite{Boyle:2019kee,Scheel:2025jct,
Healy:2017psd,Healy:2019jyf,Healy:2020vre,Healy:2022wdn,Ficarra:2026cej,
Jani:2016wkt,Ferguson:2023vta,Hamilton:2023qkv,Huerta:2019oxn}
with stored worldtube data to be
reprocessed into CCE waveform catalogs without
repeating their Cauchy evolutions.
This reprocessing is possible for both moving-puncture and
generalized-harmonic simulations, for which CCE is already in
use~\cite{Reisswig:2009us,Rashti:2024yoc,Mitman:2020pbt}.
Future runs for these catalogs can likewise yield waveforms at future null
infinity free of spurious characteristic transients.
The implications for gravitational wave science are broad and
significant.
CCE waveforms correctly capture physical features like momentum
fluxes, gravitational-wave memory, and tails~\cite{Favata:2010zu,
  Mitman:2020pbt, DaRe:2025glj, DeAmicis, Ma:2024qcv}. CCE also provides all the
Weyl components and news at future null infinity in addition to the strain.
This
enables the gravitational physics community to gain a deeper theoretical
understanding of the inspiral, merger, and ringdown of binary black holes. It
also enables improved hybridization of numerical relativity waveforms with
analytic inspiral waveforms, and the construction of
surrogate, effective-one-body, and
phenomenological models that contain these necessary physical
features.

Waveform accuracy is now limited by the outer boundary treatment in the Cauchy
simulations. However, Cauchy-characteristic matching~\cite{Ma:2023qjn} removes
that boundary, and the scheme presented here extends naturally to
it~\cite{companion}.
The implementation of Cauchy-characteristic matching in
\texttt{SpECTRE} is the subject of near-term future work, including
further improving the gauge boundary conditions that now in part
limit the accuracy of numerical relativity simulations~\cite{Sun:2026GaugeBoundary,Chaudhary:2026SpECAccuracy}.

\textit{Acknowledgments.---}
\begin{acknowledgments}
  This work was supported in part by the Sherman
  Fairchild Foundation; by NSF Grants PHY-2407742, PHY-2308615,
  PHY-2607486, OAC-2513338, and NASA award 80NSSC26K0340 at Cornell;
  by NSF Grants
  PHY-2309211, PHY-2309231, PHY-2607927, and OAC-2513339, and
  NASA award 80NSSC26K0340 at Caltech;
  and by NSF Grant PHY-2608279 and NASA award 80NSSC26K0340 at the University of
  Mississippi.
  K.M. is supported by NASA through the NASA Hubble Fellowship grant
  \#HST-HF2-51562.001-A awarded by the Space Telescope Science Institute, which is
  operated by the Association of Universities for Research in Astronomy,
  Incorporated, under NASA contract NAS5-26555. The simulations were carried
    out on the mbot cluster at Cornell and CaltechHPC at Caltech.
\end{acknowledgments}

\bibliographystyle{apsrev4-2}
\bibliography{CCEID_PRL}

\end{document}